\documentstyle[11pt,newpasp,twoside]{article}
\def\arg#1{{\it#1\/}}

\def\edcomment#1{\iffalse\marginpar{\raggedright\sl#1\/}\else\relax\fi}
\reversemarginpar

\def\be{\begin{equation}}
\def\ee{\end{equation}}
\def\bea{\begin{eqnarray}}
\def\eea{\end{eqnarray}}

\def\sm{M_\odot}
\def\etal{et al. }

\begin{document}
\title{Baryonic Dark Matter in Galaxies }
 \author{B. J. Carr}
\affil{Astronomy Unit, Queen Mary \& Westfield College, Mile End Road, London E1 4NS, UK }

\begin{abstract}
Cosmological nucleosynthesis calculations imply that many of 
the baryons in the Universe must be dark. We discuss the 
likelihood that some of these dark baryons may reside in the discs
or halos of galaxies. If they were in the form of compact objects,
they would then be natural MACHO candidates, in which case 
they are likely to be the remnants of a first generation of 
pregalactic or protogalactic Population III stars. Various candidates
have been proposed for such remnants - brown dwarfs, red dwarfs, 
white dwarfs, neutron stars or black holes - and we review the
many types of observations (including microlensing searches) 
which can be used to constrain or exclude them.
\end{abstract}

\section{Introductory Overview}

Evidence for dark matter has been claimed in many different 
contexts. There may be {\em local\/} dark matter in the Galactic disc, dark matter in the {\em halos\/} of our own and other galaxies, dark matter associated with {\em clusters\/} of galaxies and finally - if one believes that the total
cosmological density has the critical value - smoothly 
distributed {\em background\/} dark matter. Since dark matter
probably takes as many different forms as visible matter, it would be simplistic
to assume that all these dark matter problems have a single solution. The local
dark matter is almost certainly unrelated to the other ones and, while the halo and
cluster dark matter would be naturally connected in many evolutionary scenarios, there is a growing tendency to regard
the unclustered background dark matter as different from the clustered component. 

As emphasized in a recent
review by Turner (1999), there are also many different types
of dark matter candidates. The latest supernovae measurements indicate that the
cosmological expansion is accelerating, which means that the total density must be
dominated by some form of energy with negative pressure (possibly a cosmological constant). Theorists therefore now distinguish between this exotic (unclustered) dark energy and the ordinary {\em matter\/} component (with positive pressure).
The combination of the supernovae and microwave background observations suggest
that the density parameters of these two components are $\Omega_X=0.8\pm0.2$ and $\Omega_M=0.4\pm0.1$ respectively. This is compatible with the total density parameter being $1$, as expected in the inflationary scenario, but does not
definitely require this.

The matter density may itself be broken down into different components (Turner 1999). Large-scale structure observations suggest that there
must be {\em cold\/} dark matter (i.e. ``Weakly Interacting Massive Particles" or WIMPs) with a density parameter 
$\Omega_C=0.3\pm0.1$, the latest determination of the neutrino mass by the Super-Kamiokande experiment requires that there is {\em hot\/} dark matter with a density parameter $\Omega_H \approx 0.01$, 
and we will see that Big Bang nucleosynthesis calculations require that the baryonic matter (be it visible or dark) has a density parameter
$\Omega_B=0.05\pm0.005$. It is remarkable, not only that each of these components seems to be needed, but also that their densities are all 
within one or two orders of magnitude of each other:
\be
1>\Omega_X \sim \Omega_C \sim \Omega_H \sim \Omega_B >0.01.
\ee
Why this should be remains a mystery. Note that all the $\Omega$ values given above assume a Hubble parameter $H_o =65$~km~s$^{-1}$~Mpc$^{-1}$.

This paper will focus exclusively on baryonic dark matter, because that is the form
of dark matter most relevant to microlensing observations. We will address four issues: (1) What is the evidence that some of the baryons in the Universe are dark? (2) What are the reasons for believing that some of these dark baryons are in galaxies (i.e. in the form of MACHOs)? (3) What are the microlensing signatures of baryonic dark matter? (4) What constraints do these and other observations place on MACHO candidates? However,  it is important to place these considerations in the broader context discussed above.

\section{Evidence for Baryonic Dark Matter}

The main argument for both baryonic and
non-baryonic dark matter comes from Big Bang nucleosynthesis calculations. 
This is because the success of the standard picture in explaining the
primordial light element abundances only applies if the baryon 
density parameter lies in the range (Copi \etal 1995)
   \be
      0.007h^{-2} < \Omega_B < 0.022h^{-2}
   \ee
where $h$ is the Hubble parameter in units of
$100$~km~s$^{-1}$~Mpc$^{-1}$. For comparison, the latest measurements of the primordial deuterium abundance imply a much tighter constraint (Burles et al. 1999):
  \be
      0.018h^{-2} < \Omega_B < 0.020h^{-2}.
   \ee
In any case,
the upper limit implies that $\Omega_B$ is well below 1, which suggests that no
baryonic candidate could provide the matter density required by large-scale
structure observations. This conclusion also applies if one invokes
inhomogeneous nucleosynthesis since one requires $\Omega_B<0.09h^{-2}$ even
in this case (Mathews \etal 1993). On the other hand, the value 
of $\Omega_B$
allowed by eqns (2) and (3) almost certainly exceeds the density of 
visible
baryons $\Omega_V$. A careful inventory by Persic \& Salucci 
(1992)
shows that the density in galaxies and cluster gas is 
\be
\Omega_V \approx (2.2 + 0.6h^{-1.5}) \times 10^{-3} \approx 0.003
\ee
where the last approximation applies
for reasonable values of $h$. This is well below the lower limits allowed
by eqns (2) and (3), so it seems that one needs {\em both\/}
non-baryonic and baryonic dark matter.

The claim that some of the nucleosynthetic baryons must be dark is such an
important one that it is important to consider whether it can be circumvented by decreasing $\Omega_B$ or increasing
$\Omega_V$ in some way.  Both of these possibilities have been advocated.

A few years ago the anomalously high deuterium abundance measured in intergalactic Lyman-$\alpha$ clouds suggested that the
nucleosynthesis value for $\Omega_B$ could be lower than the standard one (Carswell et al. 1994, Songaila et al. 1994, Rugers \& Hogan 1996, Webb at al. 1997). However, the evidence for this has always been strongly disputed (Tytler et al. 1996). In particular, recent studies of the
quasars Q1937 and Q1009 suggest that the deuterium abundance is $3.3 \times 10^{-5}$. This corresponds to $\Omega_Bh^2=0.019$ (Burles \& Tytler 1998), which is
in the middle of the range given by eqn (3). 

The possibility that $\Omega_V$ could be larger than indicated by eqn (4) is much
harder to refute. Certainly one can now make a more precise estimate for some of the components of $\Omega_V$ than Persic \& Salucci. A recent review by 
Hogan (1999) replaces the factors in brackets in eqn (4) by 2.6 for spheroid stars, 0.86 for disc stars, 0.33 for neutral atomic gas, 0.30 for
molecular gas and 2.6 for the ionized gas in clusters. This assumes $H_o =70$~km~s$^{-1}$~Mpc$^{-1}$. However, this may still not account for all components.  For example, there may be some baryons in low surface brightness galaxies (De Blok \& McGaugh 1997) or dwarf galaxies (Loveday et al. 1997) or in a hot intergalactic medium. The last possibility is emphasized by Cen \& Ostriker (1999), who
find that - in the context of the CDM scenario - half the mass of baryons should be in warm ($10^5-10^7$K) intergalactic gas at the present epoch. 

Another possibility is that the missing baryons
could be in gas in groups of clusters. This has been emphasized by Fukugita et al. (1998), who argue 
that plasma in groups (which would be too cool to have appreciable X-ray emission)
could provide all of the cosmological nucleosynthesis shortfall. Indeed the review by Hogan (1999)
suggests that the ionized gas in groups could have a density parameter as high as $0.014$. However, it must be stressed that this estimate is based on an extrapolation of observations of rich clusters and there is no direct evidence for this.

\section{Is There Baryonic Dark Matter in Galaxies?}

Which of the dark matter problems identified in Section 1 could be baryonic? Certainly the dark matter in
galactic discs could be - indeed this is the only dark matter problem which is definitely
baryonic. Even if all discs have the 60\% dark component
envisaged for the Galaxy by Bahcall \etal (1992), this only 
corresponds to $\Omega_d \approx
0.001$, well below the nucleosynthesis bound. However, the Bahcall et al. claim is strongly disputed by Kuijken \& Gilmore (1989) and Flynn \& Fuchs (1995). Indeed recent Hipparcos observations suggest that the dark disc fraction is below 10\% (Cr\'ez\'e \etal 1998).

The more interesting question is whether the 
halo dark matter could be baryonic. If the Milky Way is typical, the density associated with halos would 
be $\Omega_h \approx 0.03h^{-1}(R_h/100\mbox{kpc})$, where $R_h$ is the (uncertain) halo radius, so the upper limit in eqn (2) 
implies that {\em all\/} the dark matter in halos could be baryonic 
providing
$R_h<70h^{-1}$kpc. This is marginally possible for our galaxy (Fich \& 
Tremaine 1991), in which case the dark baryons could be contained in the remnants of a first generation of ``Population III" stars (Carr 1994). This corresponds to the ``Massive Compact Halo Object" or ``MACHO" scenario and has attracted considerable interest as a result of the LMC microlensing observations. Even if halos are larger than $70h^{-1}$kpc, and studies of the kinematics of other spirals by Zaritsky et al. (1993) suggest that they could be as large as $200h^{-1}$kpc, a considerable fraction of their mass could still be in stellar remnants.

Probably the only direct evidence that there are dark baryons in galaxies comes from
studying the density profiles and rotation curves in dwarf galaxies. It is well known that the presence of
dark matter in dwarf galaxies requires that it cannot be entirely hot. However,
if the dark matter consisted only of WIMPs, one would expect it to have the standard 
Navarro-Frenk-White density profile (Navarro et al. 1997) and Burkert  \& Silk (1999) claim that the profile for
DDO 154 is very different from this. They argue that the measurements indicate the presence of a centrally condensed baryonic dark component, having about 25\% of the total dark matter density. Another possible signature of a baryonic halo would be {\it flattening} of the halo since more dissipation would 
be expected in this case. For our galaxy observations are best fit by an axis ratio $q\approx 0.6$, which
constrains the fraction of the halo in MACHOs but does not
necesssarily require them (Samurovic et al. 1999).
 
On theoretial grounds one would {\it expect} the halo dark matter to be a mixture of WIMPs and MACHOs. For since the cluster dark matter must be predominantly cold, one would expect at least some of it to clump into galactic
halos. The relative densities of the two components would depend sensitively on
the formation epoch of the Population III stars. If they formed pregalactically, one would expect
the halo ratio to reflect the cosmological background ratio (viz. $\Omega_B/\Omega_C \approx 0.1$). However, if they formed protogalactically, the ratio could be larger
since the baryons could have dissipated and become more centrally concentrated 
than the WIMPs.
 
In order to distinguish between the pregalactic and protogalactic scenarios, it is important to gain independence evidence
about the formation epoch of the putative MACHOs. At moderate redshifts one can obtain a lower limit to the baryon density by studying Lyman-$\alpha$ clouds. The simulations of Weinberg et al. (1997) suggest that the
density parameter of the clouds must be at least $0.017h^{-2}$ at a redshift of 2 and this is already
close to the upper limit given by eqn (3). By today some of these baryons might have been transformed into a hot intergalactic medium or stars but this suggests that there was little room for any dark baryons before $z=2$. On the other hand, we will see
in Section 5.4 that background light constraints require that any massive Population III stars must have formed much earlier than this.

\section{Lensing Constraints}

One of the most useful signatures of compact
objects is their gravitational lensing effects. Indeed it is
remarkable that lensing could permit their detection over the 
entire mass range $10^{-7}\sm$ to $10^{12}\sm$.
There are two distinct lensing effects and these probe different 
but nearly overlapping mass ranges: macrolensing (the multiple-imaging of a
source) can be used to search for objects larger than $10^5\sm$,
while microlensing (modifications to the intensity of a source) can 
be used for objects smaller than $10^3\sm$. The current constraints 
on the density parameter of compact objects in various mass ranges 
are brought together in Figure 1, shaded regions being excluded. 
Although the focus of this meeting is mainly
on microlensing, it is useful to bring all the limits together. Most of these
limits presume a cosmological distribution of lenses but they are still
applicable for lenses confined to halos provided the halos cover the sky.  
\vspace{10pt}

{\em Macrolensing by Compact Objects}. If one has a population of
compact objects with mass $M$ and density parameter 
$\Omega_c$, then
the probability of one of them multiply-imaging a source at 
a cosmological redshift
and the separation between the images are roughly
   \begin{equation}
      P \approx 0.1 \, \Omega_c, \quad \quad \quad \theta 
\approx
      6 \times 10^{-6}(M/\sm)^{1/2} h^{1/2}~\mbox{arcsec}
   \end{equation}
(Press \& Gunn 1973). One can therefore use upper limits on the
frequency of macrolensing for different image separations to 
constrain
$\Omega_c$ as a function of $M$. In particular, 
in the context of quasars, VLA
observations imply $\Omega_c(10^{11}-10^{13}\sm)<0.4$ (Hewitt 
1986)
and HST data imply
 $\Omega_c(10^{10}-10^{12}\sm)<0.02$ (Surdej \etal 1993). To 
probe
smaller scales, one can use high resolution radio sources: Kassiola 
et
al. (1991) have invoked lack of lensing in 40 VLBI objects to 
infer
$\Omega_c(10^7- 10^9\sm)<0.4$, while a study of VLBA sources
leads to a limit $\Omega_c(10^6- 10^8\sm)<0.03$ (Henstock 1996). 
Other techniques (eg. speckle interferometry) could 
strengthen these
constraints, as indicated by the broken lines in Figure 1. It should be stressed
that the expression for $P$ in eqn (5) has some dependence on the cosmological model. Indeed one of the important recent uses of macrolensing searches is to constrain
the cosmological constant, the observations requiring $\Omega_{\Lambda}<0.7$.
\vspace{10pt}

{\em Microlensing in Macrolensed Sources}. If a galaxy is suitably positioned to 
image-double a quasar, then there is also a high probability that 
an individual halo object will traverse the line of sight of one of the
images and this will give intensity fluctuations in one
but not both images (Gott 1981). The effect would be observable for objects 
bigger
than $10^{-4}\sm$ but the timescale of the fluctuations
$\sim 40(M/\sm)^{1/2}$y would make them detectable only for  
$M<0.1~\sm$. There is already evidence of this effect for the 
quasar
2237+0305 (Irwin \etal 1989), the observed
timescale for the variation in the luminosity of one of the images
possibly indicating a mass below $0.1~\sm$ (Webster \etal 1991). However, because
the optical depth is high, the mass estimate is very uncertain and
a more recent analysis suggests that it could be in the range
$0.1 - 10~\sm$ (Lewis et al. 1998), in which case the lens could be an
ordinary star. The absence of this effect in the quasar 0957+561 has also been used to exclude MACHOs with mass in the range $10^{-7} - 10^{-3}\sm$ from making up all of the halo of the intervening galaxy, although the precise limit has some dependence on the quasar size (Schmidt \& Wambsganss 1998). Another application of this method
is to seek microlensing in a compact radio source which is macrolensed by a galaxy. Indeed Koopmans \& Bruyn (2000) claim to have detected this effect for the 
CLASS gravitational lens B1600+434. The inferred lens mass is around $0.5~\sm$, comparable to the mass implied by the LMC data. This result is discussed in more detail at this meeting.
\vspace{10pt}

{\em Microlensing of Quasars}. More dramatic 
but rather controversial evidence for the microlensing of quasars 
comes from Hawkins (1993, 1996, 1999), who has been monitoring 300 quasars in 
the redshift range $1-3$ for nearly 20 years using a wide-field 
Schmidt camera. He finds quasi-sinusoidal variations with an amplitude of 
0.5 magnitudes on a timescale 5~y and attributes this to lenses with 
mass $\sim 10^{-3}\sm$. The crucial point is that the timescale 
decreases with
increasing redshift, which is the opposite to what one would 
expect for intrinsic variations, although this has been disputed (Alexander 1995, 
Baganoff \& Malkan 1995). The timescale also increases with the luminosity of the quasar and he explains this by noting that the
variability timescale should scale with the size of the accretion 
disc (which should itself correlate with luminosity). A rather worrying
feature of Hawkins' claim is that he requires the density of the
lenses to be close to critical (in order that the sources are
transited continuously), so he has to invoke primordial black 
holes which form at the quark-hadron phase transition 
(Crawford \& Schramm 1982). 
However, this requires fine-tuning since the fraction of the
Universe going into black holes at this transition must only be about $10^{-9}$. As discussed in Section 5.5, Walker (1999) has proposed that Hawkins' lenses might also be jupiter-mass gas clouds.  
\vspace{10pt}

{\em Line-Continuum Effects for Quasars}. In some circumstances, 
the continuum
part of the quasar emission will be microlensed but not the line part. This is
because only the continuum region may  be small enough to act as a point-like source.
 (For a lens at a cosmological distance the Einstein
radius is $0.05(M/\sm)^{1/2}h$~pc, whereas the sizes of the optical
continuum and line regions are of order $10^{-4}$pc and 1~pc
respectively.) This would decrease the equivalent width of the emission lines, so in statistical studies of 
many quasars one would expect the characteristic equivalent width of
quasar emission lines to decrease as one goes to higher redshift
because there would be an increasing probability of having an
intervening lens. Dalcanton \etal (1994) compared the
equivalent widths for a high and low redshift sample of quasars 
and
found no difference. They inferred
 \begin{equation}
      \Omega_c(0.001-60~\sm)<0.2, \quad
      \Omega_c(60-300~\sm)<1, \quad
\Omega_c(0.01-20~\sm)
      <0.1.
   \end{equation}
The mass limits come from the fact that the amplification of even 
the continuum region would be unimportant for $M<0.001~\sm$, 
while the
amplification of the line regions would be important (cancelling 
the effect) for $M>20~\sm$ if $\Omega_c=0.1$ or $M>60~\sm$ if
$\Omega_c=0.2$ or $M>300~\sm$ if $\Omega_c=1$. These limits 
are indicated in Fig.1 and are marginally incompatible with 
Hawkins' claim that $\Omega_c(10^{-3}\sm) \sim 1$.
\vspace{10pt}

\begin{figure}\label{F1}
\vskip 3.3in
\caption{Lensing constraints on density parameter for 
compact objects}
\end{figure}

{\em Lensing of Quasars by Dark Objects in Clusters}.
If halos are made of MACHOs,  one would also expect some of these to be spread throughout a cluster of galaxies. This is because individual galaxies
should be stripped of some of their outer halo as a result of collisions and tidal interactions. This method is sensitive to MACHOs in the
mass range $10^{-6} - 10^{-3}~\sm$. Tadros et al. (1998) have therefore looked for the microlensing of quasars by MACHOs in the Virgo
cluster: four months of observations of 600 quasars with the UK Schmidt telescope have yielded no candidates and this already implies that less than half
the mass of Virgo is in $10^{-5}~\sm$ objects. A more extensive follow-up
campaign is currently underway. This technique would also be
sensitive to cold molecular clouds of the kind advocated by Walker (1999). 
\vspace{10pt}

{\em Microlensing of Supernovae by Halos}.
In principle galactic halos could produce luminosity variations
in high redshift supernovae, many of which are now routinely detected as
a result of supernova searches. As pointed out by Metcalf \& Silk (1999), a
particularly interesting aspect of this effect is that it could discriminate between
what they term ``macroscopic" and ``microscopic" dark matter,
corresponding to MACHO and WIMP halos respectively. This is because the distribution of amplifications would be different in these two cases,
although the shape of the distribution is also sensitive to the cosmological and halo
model.
\vspace{10pt}

{\em Microlensing of Stars in LMC}. 
Attempts to detect microlensing by objects in our own halo by 
looking for
intensity variations in stars in the Magellanic Clouds and the
Galactic Bulge have now been underway for a decade (Paczynski 1996). This method
is sensitive to lens masses in the range $10^{-7}-10^2\sm$ but the probability of an
individual star being lensed is only $\tau \sim 10^{-6}$, so one 
has to
look at many stars for a long time (Paczynski 1986). The duration and likely 
event rate are
\be
P\sim 0.2(M/\sm)^{1/2}y,\quad \Gamma \sim N \tau P^{-1} \sim (M/\sm)^{-1/2}y^{-1}
\ee
where $N\sim 10^6$ is the number of stars.
As discussed elsewhere in this meeting, the MACHO 
group currently has 13-17 LMC events and the durations span the range 
$34-230$ days (Alcock et al. 
2000). For a standard halo model, the data suggest an average lens mass of around $0.5~\sm$ and a halo fraction of 0.2, with the 95\% confidence ranges being $0.15-0.9~\sm$ and $0.08-0.5$. The mass is comparable with the earlier estimates but the
fraction is somewhat smaller (Alcock et al. 1997). This
might appear to indicate that the MACHOs are white
dwarfs but, as discussed in Section 5.2, this would seem to be excluded on astrophysical grounds, so this presents a dilemma for MACHO enthusiasts. 
One possible resolution is to invoke a less conventional candidate; for
example, primordial black holes forming at the
quark-hadron phase transition might have the required mass 
(Jedamzik 1997) and the microlensing implications of this scenario have been
studied by Green (1999). Perhaps the
most important result of the LMC searches is that they {\em eliminate} many candidates. Indeed the combination of the MACHO and EROS results already excludes objects in the mass range $5\times10^{-7}-0.002~\sm$ from having more than 0.2 of the halo density (Alfonso et al. 1997).
\vspace{10pt}

{\em Microlensing of Stars in M31}.
The LMC studies are complemented by searches for microlensing of stars in M31.
In this case, the sources are too distant to resolve individually (i.e. there are many
stars per pixel), so a lensing event
is observed only if the amplification is large enough for the source to stand out 
above the background, but observations of
the LMC already demonstrate the efficacy of the method (Melchior et al. 1999). For sources in M31 the halo objects may reside in our own galaxy or M31 but  the crucial point
is that one expects an asymmetry between the far and near side of the disc. Two
groups have been involved in this work: the AGAPE collaboration (Ansari et al. 1997),
who use the ``pixel" method, and the VATT-Columbia collaboration, who use
``differential image photometry" (Crotts \& Tomaney 1997). The AGAPE team have been monitoring seven fields
in M31 in red and blue and have already detected one good lensing candidate 
(Ansari et al. 1999). The important theoretical implications of this approach are considered by Kerins et al. (2000) and this is discussed further at this meeting.

\section{Constraints on MACHO candidates}

Although one cannot state definitely that MACHOs exist, one can already place 
important constraints on the possible candidates. In this section we will discuss
each candidate in turn, focussing particularly  on brown dwarfs, red dwarfs, white dwarfs and black holes. The combined constraints, including the microlensing ones discussed above,  are indicated in Figure 2. This
shows which candidates are excluded by various types of observational signature.
A cross indicates that exclusion is definite, while a question mark indicates that it is tentative. Candidates associated with one or more crosses should clearly be rejected but
those with question marks alone may still be viable.
Although no candidate is entirely free of crosses or questions marks, the title of a
recent paper by Freese \etal (1999), ``Death of Stellar Baryonic Dark Matter", suggesting that there are no viable MACHO candidates, may be overly pessimistic.

\begin{figure}\label{F2}
\vskip 3.0in
\caption{Constraints on and exclusions of MACHO candidates}
\end{figure}

\subsection{Brown Dwarfs}

Objects in the range $0.001-0.08~\sm$ would never burn hydrogen and are termed ``brown dwarfs" (BDs). They represent a balance between gravity and degeneracy pressure. Objects below $0.001~\sm$, being held together by intermolecular rather than gravitational forces, have atomic density and are sometimes termed ``snowballs" (SBs). However, such objects would have evaporated within the age of the Universe if they were smaller than $10^{-8}\sm$ (De Rujula et al. 1992) and there are various encounter constraints for snowballs larger than this (Hills 1986). 

It has been argued that objects below the hydrogen-burning 
limit  may form efficiently in pregalactic or protogalactic cooling flows (Ashman \& Carr 1990, Thomas \& Fabian 1990) but the direct evidence for such objects
remains weak. While some BDs have been found as
companions to ordinary stars, these can only have a tiny cosmological density and 
it is much harder to find isolated field BDs. The best argument therefore comes from 
extrapolating the initial mass function (IMF) of hydrogen-burning stars to lower masses than can be observed directly. The IMF for Population I stars ($dN/dm \sim m^{-\alpha}$ with $\alpha<1.8$) suggests that only 1\% of the disc could be in BDs (Kroupa et al. 1995). However, one might wonder whether $\alpha$ could be larger, increasing the BD fraction, for zero-metallicity stars. Although there are theoretical reasons for entertaining this possibility, earlier observational claims that low metallicity  objects have a steeper IMF than usual are now discredited. Indeed observations of 
Galactic and LMC globular clusters (Elson et al. 1999) and dwarf spheroidal field stars
(Feltzing et al. 1999) suggest that the IMF is {\it universal} with $\alpha <1.5$ at low masses (Gilmore 1999). This implies that the BD fraction is much less than 1\% by mass. However, it should be stressed that nobody has yet measured the IMF in the sites which are most likely to be associated with Population III stars.

We have seen that the LMC microlensing results would now seem to exclude a large fraction of BDs in our own halo. Although Honma \& Kan-ya (1998) have presented 100\% BD models, these would require falling rotation curves and most theorists would regard these as rather implausible.  Another exotic possibility, suggested by Hansen (1999), is ``Beige Dwarfs" in the
mass range  $0.1-0.3~\sm$. Such objects are larger than the traditional BD upper limit but they are supposed to
form by sufficiently slow accretion that they never ignite their nuclear fuel.

\subsection{Red Dwarfs}

Discrete source counts for our own Galaxy suggest that the fraction of the halo mass in low mass hydrogen burning stars - red dwarf (RDs) - must be less than 1\% (Bahcall et al. 1995, Gould et al. 1998, Freese et al. 1999). These limits might be weakened if the stars were clustered
(Kerins 1997) but not by much. For other galaxies, the best constraint on the red dwarf fraction comes from upper limits on the halo red light and such studies go back 
several decades (Boughn \& Saulson 1983). 

The discovery of red light around NGC 5907 by Sackett et al. (1994), apparently emanating from low mass stars with a density profile like that of the halo, was therefore a particularly interesting development. This detection was confirmed in V and I by Lequeux et al. (1996) and in J and K by James \& Casali (1996). However, the suggestion that
the stars might be of primordial origin (with low metallicity) was contradicted by the results of Rudy et al. (1997), who found that the color was indicative of low mass stars with solar metallicity.
Furthermore, it must be stressed that the red light light has only been observed within a few kpc and no NIR emission is detected at 10-30 kpc (Yost et al. 1999). Both these points go against the suggestion that the red light is associated with MACHOs.

Recently it has been suggested that the red light seen
in NGC 5907 is more likely to derive from a disrupted dwarf galaxy, the stars of which
would naturally follow the dark matter profile (Lequeux et al. 1998), or to be a ring left over from a disrupted dwarf spheroidal galaxy (Zheng et al. 1999). However, in this case one would expect of order a hundred bright giants for a standard IMF, whereas NICMOS observations find only one (Zepf et al. 1999). This requires that either the galaxy is much further away than expected (24 Mpc) or it has a very low metallicity or the dwarf-to-giant ratio is very large (requiring a very steep IMF with $\alpha>3$). There is clearly still a mystery here. In any case, NGC 5907 does not seem to be typical since ISO observations of four other edge-on bulgeless spiral galaxies give no evidence for red halos (Gilmore \& Unavane 1998).

\subsection{White Dwarfs}

A few years ago white dwarfs (WDs) were regarded as rather implausible dark matter candidates. One required a very contrived IMF, lying between $2~\sm$ and $8~\sm$, 
in order to avoid excessive production of light or metals (Ryu et al. 1990); the fraction
of WD precursors in binaries would be expected to produce too many type 1A supernovae (Smecker \& Wyse 1991); and the halo fraction was constrained to be less than 10\% in order to avoid the luminous precursors contradicting the upper limits from galaxy counts (Charlot \& Silk 1995).  The observed WD luminosity function
also placed a severe lower limit on the age of any WDs in our own halo (Tamanaha et al. 1990). 

More recent constraints have strengthened these limits. A study of CNO production suggests that a halo comprised entirely of WDs would overproduce C and N compared to O by factors as
large as 60 (Gibson \& Mould 1997) and,
although one might be able to circumvent this constraint in some circumstances,  a similar limit comes from considering helium and deuterium production (Field et al. 2000). Extragalactic
background light limits now require that the halo WD fraction be less than 6\% (Madau \& Pozzetti 1999) and 
the detection of TEV $\gamma$-rays from the the galaxy Makarian 501 (which 
indirectly constrains the infrared background) requires that the WD density parameter
be less than $0.002h^{-1}$ (Graff et al. 1999).

The ``many nails in the coffin"  of the WD scenario are confounded by the results of the LMC microlensing observations, the lens mass estimate for
which suggests that WDs are the most plausible explanation. Not surprisingly, therefore, theorists have been trying to resuscitate the scenario. At least some of the afore-mentioned limits must be reconsidered in view of recent claims by
Hansen (1998) that metal-poor old WDs with hydrogen envelopes could be much bluer and brighter than previously supposed, essentially because the light emerges from deeper in the atmosphere. 

This suggestion has been supported by HST observations of Ibata et al. (1999), who claim to have detected five candidates of this kind. The objects are blue and isolated
and show high proper motion. They infer that they are $0.5~\sm$ hydrogen-atmosphere WDs with an age of around $12$ Gyr. Three such objects have now been 
identified spectroscopically (Hodgkin et al. 2000, Ibata et al. 2000), so this possibility
must be taken very seriously. However, this does not circumvent the nucleosynthetic  arguments against WDs. 

\subsection{Black Hole Remnants}

Stars bigger than $8~\sm$ would leave neutron star (NS) remnants, while
those in some range above about $20~\sm$ would leave black hole (BH) remnants. However, neither of these would be plausible candidates for either the disc or halo dark matter because their precursors would have unacceptable nucleosynthetic yields. 
Stars larger than $200~\sm$ are termed ``Very Massive Objects" or VMOs and might 
collapse to black holes without these nucleosynthetic consequences (Carr et al. 1984). However, during their main-sequence phase, such VMOs would be expected to generate of a lot of background light. By today this should have been shifted into the infrared or submillimetre band, as a result of either redshift effects or dust reprocessing, so one would expect a sizeable infrared/submillimetre cosmic background (Bond et al. 1991). Over the last few decades there have been 
several reported detections of such a background but these have usually turned out to be false alarms. COBE does now seem to have detected a
genuine infrared background (Fixsen et al. 1998) but this can probably be attributed to 
ordinary Population I and II stars. In any case, the current constraints on such a background strongly limit the density of any VMOs unless they form at a very high redshift.  For this reason massive Population III stars would need to be pregalactic rather than protogalactic. 

Stars larger than $10^5~\sm$ - termed ``Supermassive Objects" or SMOs - would collapse directly to black holes without any nucleosynthetic or background light 
production. However, supermassive black holes would still have noticeable
lensing effects, as discussed in Section 4, and dynamical effects. The latter have been investigated  by
many authors and are reviewed in detail by Carr \& Sakellariadou (1998). The constraints on black holes in our own disc - due to the disruption of open clusters - and in our own halo - due to the heating of disc stars, the disruption of globular clusters and dynamical friction effects - are 
indicated by the shaded regions in Figure 3. Although it has been claimed that there is positive evidence for some of these effects, such as disc heating (Lacey \& Ostriker 1985), the interpretation of this evidence is not clear-cut. 
It is therefore more natural to regard these dynamical effects  as merely imposing an upper limit on the density of black holes or indeed any other type of compact object.

The limits in Figure 3 are expressed in terms of the
density parameter, taken to be 0.001 for the disc and 0.1 for the halo. The figure also shows the dynamical constraints for black holes in clusters of galaxies or intergalactic space, although this goes beyond the context of the present discussion. It should be
stressed that many of these limits would also apply if the black holes were replaced by ``dark clusters"
of smaller objects, a scenario which has been explored by many authors (Carr \& Lacey 1987, Ashman 1990, Kerins \& Carr 1994, De Paolis et al. 1995, Moore \& Silk 1995).

\begin{figure}\label{F3}
\vskip 3.2in
\caption{Dynamical constraints on the density parameter for black holes of
mass $M$ located in the Galactic disc, the Galactic halo, clusters of galaxies and intergalactic space}
\end{figure}

\subsection{Cold Clouds}

The suggestion that the halo dark matter could be in cold clouds was first made by 
Pfenniger et al. (1994). They envisaged the clouds having a mass of around $10^{-3}~\sm$
and being distributed in a disc, which now seems dynamically rather implausible, but
several people have proposed a similar scenario with a spheroidal halo of clouds (De Paolis et al. 1995).
Walker (1999) argues that such clouds could explain both the ``Extreme Scattering Events" detected by radio observations in our own galaxy and the quasar
microlening events claimed by Hawkins. Indeed Walker \& Wardle (1999) advocate a model in which the halo entirely comprises such clouds, with visible stars being formed as a
result of collisions between these clouds. They claim that this scenario naturally produces various observed features of galaxies. Although clouds of  $10^{-3}~\sm$ could not
explain the LMC microlensing events, Draine (1998) argues that such clouds
might still produce the apparent microlensing events through gas absorption effects. The interaction of cosmic rays with such clouds would also produce an interesting
$\gamma$-ray signature (Sciama 2000, De Paolis et al. 1999). 

\section{Conclusions}

Although it is premature to assess the importance of baryonic dark matter definitively, there have been many interesting developments in this field in the last few years and various conclusions can be drawn. 

1) Although cosmological nucleosynthesis calculations suggest that many baryons are dark, one cannot be  sure that the dark baryons are inside galaxies. The more
conservative conclusion would be that they are contained in an intergalactic medium or in gas within groups or clusters of galaxies.

2) Over the years there have been several observational claims of effects which seem to indicate the existence of MACHOs but these have usually turned out to be false alarms.
For example, the discovery of a red halo around NGC 5907 is suggestive but we have seen that its interpretation is far from clear.

3) Currently the only {\it positive} evidence for MACHOs comes from microlensing observations. The LMC results suggest that white dwarfs may be the best MACHO candidate but it must be stressed that the mass estimate upon which this inference is based is sensitive to assumptions about the halo model. In any case, the large number of arguments which have been voiced against white dwarfs in the past cannot be brushed aside too cavalierly. The detection of microlensing in a compact radio source also gives a mass in the white dwarf range but the Hawkins result requires a much smaller mass. It would perhaps be strange to have two distinct populations.

4) We have seen that there are many important {\it constraints} on MACHO candidates, not only from microlensing but also from a wide variety of other astrophysical effects. These constraints are summarized in Figure 2. Until there is a definite detection, therefore, the best strategy is to proceed by {\it eliminating} candidates, on the Sherlock Holmes principle that whatever candidate remains, however implausible, must be correct.

5) What is clearly missing from current speculation is a good cosmological scenario for the formation of the MACHOs. There is considerable uncertainty as to whether they form pregalactically
(as suggested by background light constraints) or more recently (as suggested by 
observations of Lyman-$\alpha$ clouds). There is also ambiguity as whether they comprise a thick disc, as proposed by Gates \& Gyuk (2000), or a spheroidal halo.

\end{document}